\documentclass[10pt,conference,letterpaper]{IEEEtran}
\usepackage{times,amsmath,amssymb}
\usepackage{graphicx}
\usepackage{makeidx}
\usepackage{multirow}
\usepackage{amssymb}
\usepackage{latexsym}
\newtheorem{theorem}{Theorem}

\newcommand{\qed}{\hfill \ensuremath{\Box}}
\long\def\symbolfootnote[#1]#2{\begingroup%
\def\thefootnote{\fnsymbol{footnote}}\footnote[#1]{#2}\endgroup}
\title{Fully Dynamic Data Structure\\ for Top-$k$ Queries on Uncertain Data}
\author{%
{Manish Patil{\small $~^{1}$}, Rahul Shah{\small $~^{2}$}, Sharma V. Thankachan{\small $~^{3}$} }%
\vspace{1.6mm}\\
\fontsize{10}{10}\selectfont\itshape
$~^{}$Computer Science Department, Louisiana State University,\\
Baton Rouge, LA, USA\\
\fontsize{9}{9}\selectfont\ttfamily\upshape
$~^{1}$mpatil@csc.lsu.edu\\
$~^{2}$rahul@csc.lsu.edu\\
$~^{3}$thanks@csc.lsu.edu%
}

\begin{document}
	
\maketitle

\begin{abstract} 
	

Top-$k$ queries allow end-users to focus on the most important (top-$k$) answers amongst those which satisfy the query. In traditional databases, a user defined score function assigns a score value to each tuple and a top-$k$ query returns $k$ tuples with the highest score. In uncertain database, top-$k$ answer depends not only on the scores but also on the membership probabilities of tuples. Several top-$k$ definitions covering different aspects of score-probability interplay have been proposed in recent past~\cite{R10,R4,R2,R8}. Most of the existing work in this research field is focused on developing efficient algorithms for answering top-$k$ queries on static uncertain data. Any change (insertion, deletion of a tuple or change in membership probability, score of a tuple) in underlying data forces re-computation of query answers. Such re-computations are not practical considering the dynamic nature of data in many applications. In this paper, we propose a fully dynamic data structure that uses ranking function $PRF^e(\alpha)$ proposed by Li et al.~\cite{R8} under the generally adopted model of $x$-relations~\cite{R11}. $PRF^e$ can effectively approximate various other top-$k$ definitions on uncertain data based on the value of parameter $\alpha$. An $x$-relation consists of a number of $x$-tuples, where $x$-tuple is a set of mutually exclusive tuples (up to a constant number) called alternatives. Each $x$-tuple in a relation randomly instantiates into one tuple from its alternatives. For an uncertain relation with $N$ tuples, our structure can answer top-$k$ queries in $O(k\log N)$ time, handles an update in $O(\log N)$ time and takes $O(N)$ space. Finally, we evaluate practical efficiency of our structure on both synthetic and real data.

\end{abstract}


\section{Introduction}
The efficient processing of  uncertain data is an important issue in many application domains because of the imprecise nature of data they generate. The nature of uncertainty in data is quite varied, and often depends on the application domain. In response to this need, much effort has been devoted to modeling uncertain data~\cite{R11,R3,R1,R7,R9}. Most  models have  been adopted to possible world semantics, where an uncertain relation is viewed as a set of possible instances (worlds) and correlation among the tuples governs generation of these worlds.

Consider traffic monitoring application data~\cite{R10} (with modified probabilities) as shown in Table~\ref{tab_data}, where radar is used to detect car speeds. In this application, data is inherently uncertain because of errors in reading introduced by nearby high voltage lines,  interference from near by car, human operator error etc. If two radars at different locations detect the presence of the same car within a short time interval, such as tuples $t_2$ and $t_4$ as well as $t_3$ and $t_6$, then at most one radar reading can be correct. We use $x$-relation model to capture such corrections. An $x$-tuple $\tau$ specifies a set of exclusive tuples, subject to the constraint $\sum_{t_i\in \tau} Pr(t_i) \leq 1$. The fact that $t_2$ and $t_4$ cannot be true at the same time, is captured by the  $x$-tuple $\tau_1 = \{t_2, t_4\}$. Similarly $\tau_2= \{t_3, t_6\}$. Probability of a possible world is computed based on the existence probabilities of tuples present in a world and absence probabilities of tuples in the database that are not part of a possible world. For example, consider the possible world $pw = \{t_1, t_2, t_3\}$. Its probability is computed by assuming the existence of $t_1$, $t_2$, $t_3$, and the absence of $t_4$, $t_5$, and $t_6$. However since $t_2$ and $t_4$ are mutually exclusive presence of tuple $t_2$ implies absence of $t_4$ and same is applicable for tuples $t_3$ and $t_6$. Therefore, $Pr(pw) = 0.3 \times 0.4 \times 0.2 \times (1-0.3) = 0.0168$.

\begin{table}[!t]
\centering

   \caption{Traffic monitoring data: $t_1$ ,$\{t_2, t_4\}$, $\{t_3, t_6\}$, $t_5$}     
   \label{tab_data}

   \begin{small}
   \begin{tabular}{|l|l|l|l|l|l|}
   \hline
       Time & Car  & Plate & Speed & Prob & Tuple \\
        & Loc  & No &  &  & Id \\ \hline
        11:55 & L1 & Y-245 & 130 & 0.30 & $t_1$ \\ \hline
        11:40 & L2 & X-123 & 120 & 0.40 & $t_2$ \\ \hline
       12:05 & L3 & Z-541 & 110 & 0.20 & $t_3$ \\ \hline
       12:15 & L4 & X-123 & 105 & 0.50 & $t_4$ \\ \hline
       12:10 & L5 & L-110 & 95 & 0.30 & $t_5$ \\ \hline
       11:35 & L6 & Z-541 & 80 & 0.45 & $t_6$ \\ \hline
   \end{tabular}
   \end{small}
\end{table}

Top-$k$ queries on a traditional certain database have been well studied. For such cases, each tuple is associated with a single score value assigned to it by a scoring function. There is a clear total ordering among tuples based on score, from which the top-$k$ tuples can be retrieved. However, for answering a top-$k$ query on uncertain data, we have to take into account both, ordering based on scores and ordering based on existence probabilities of tuples. Depending on how these two orderings are combined, various top-$k$ definitions with different semantics have been proposed in recent times. Most of the existing work studies only the problem of answering a top-$k$ query on a static uncertain data. Though the query time of an algorithm depends on the choice of a top-$k$ definition, linear scan of tuples achieves the best bound so far. Therefore, recomputing top-$k$ answers in an application with frequent insertions and deletions can be extremely inefficient. In this paper, we present a fully dynamic structure of size $O(N)$ that always maintains the correct answer to the top-$k$ query for an uncertain database. The structure is based on a decomposition of the problem so that updates can be handled efficiently. Our structure can answer the top-$k$ query in $O(k\log N)$ time, handle update in $O(\log N)$ time.\\

\emph{Outline:} In Section~\ref{S2} we review different top-$k$ definitions proposed so far and try to compare them against a parameterized ranking function $PRF^e(\alpha)$ proposed by Li et al.~\cite{R8}. We choose $PRF^e(\alpha)$ over other definitions as it can approximate many of the other top-$k$ definitions and can handle data updates efficiently. After formally defining the problem (Section~\ref{S3}), we explain how  $PRF^e(\alpha)$ can be computed using divide and conquer approach (Section~\ref{S4}), which forms the basis of our data structure explained in Section~\ref{S5}. We present experimental study with real and synthetic data sets in Section~\ref{S6}. Finally we review the related work in Section~\ref{S7} before concluding the paper.

\section{Top-$k$ queries on uncertain data}
\label{S2}
Soliman et al.~\cite{R10} first considered the problem of ranking tuples when there is both a score and probability for each tuple. Several other definitions of ranking have been proposed since then for probabilistic data. 
\begin{itemize}
\item Uncertain Top-$k$ (U-Topk)~\cite{R10}: It returns a $k$-tuple set that appears as top-$k$ answer in possible worlds with maximum probability.

\item Uncertain Rank-$k$ (U-Ranks)~\cite{R10}: It returns a tuple for each $i$, such that it has maximum probability of appearing at rank $i$ across all possible worlds.

\item Probabilistic Threshold Query (PT-k)~\cite{R4}: It returns all the tuples with probability of appearing in top-$k$ greater than a user specified threshold.

\item Expected Rank (E-Rank)~\cite{R2}: $k$ tuples with highest value of expected rank (er($t_i$)) are returned.

\begin{equation*} 
	\begin{split}
		er(t_i) = \sum Pr(pw) rank_{pw}(t_i)
	\end{split} 
\end{equation*}
where $rank_{pw}(t_i)$ denotes rank of $t_i$ in a possible world $pw$. In case $t_i$ does not appear in possible world, $rank_{pw}(t_i)$ is defined as $|pw|$.

\item Expected Score (E-Score)~\cite{R2}: $k$ tuples with highest value of expected score (es($t_i$)) are returned.

\begin{equation*} 
	\begin{split}
		es(t_i) = Pr(t_i) score(t_i)
	\end{split} 
\end{equation*}

\item Parameterized Ranking Function (PRF)~\cite{R8}: $PRF$ in its most general form is defined as, 
\begin{equation} 
	\label{eq_prfwdef}
	\begin{split}
		\Upsilon(t_i)& = \sum_{r} w(t_i, r) \times Pr(t_i, r)
	\end{split} 
\end{equation}

where $w$ is the weight function that maps a given  tuple-rank pair to a complex number and $Pr(t_i, r)$ denotes the probability of a tuple $t_i$ being ranked at position $r$ across all possible worlds. A top-$k$ query returns those $k$ tuples with the highest $\Upsilon$ values. Different weight functions can be plugged in to the above definition to get a range of ranking functions, subsuming most of top-$k$ definitions listed above. A special ranking function $PRF^e (\alpha)$ is obtained by choosing $w(t_i, r) = \alpha^{r-1}$, where $\alpha$ is a constant. Experimental study in~\cite{R8} reveals that for some value of $\alpha$ with the constraint $\alpha < 1$, $PRF^e$ can approximate many existing top-$k$ definitions.

\end{itemize}

Algorithms for computing top-$k$ answers using the above ranking functions have been studied for static data. Any changes in the underlying data forces re-computation of query answers. To understand the impact of a change on top-$k$ answers, we analyze relative ordering of the tuples before and after a change, based on these ranking functions. 

Let $T = t_1, t_2, .., t_N$ denote independent tuples sorted in non-increasing order of their score. We choose insertion of a tuple as a representative case for changes in $T$, and monitor its impact on relative ordering of a pair of tuples ($t_i$, $t_j$). Since {\tt E-Score} of a tuple depends only on its score and existence probability, ordering is preserved for all ($t_i$, $t_j$) pairs in $T$. For ranking functions {\tt U-Ranks, PT-k} ordering of tuples ($t_i$, $t_j$) may or may not be preserved by insertion and cannot be guaranteed when the score of a new tuple is higher than that of $t_i$ and $t_j$. Hence, existing top-$k$ answers do not provide any useful information for re-computation of query answers.  {\tt E-Rank} further complicates the matter as expected rank of a tuple depends on both higher and lower scored tuples. However, when tuples are ranked using $PRF^e(\alpha)$, the scope of disturbance in the relative ordering of tuples is limited as explained in later sections. This enables efficient handling of updates in the database. Therefore, this ranking function is well suited for answering top-$k$ queries on a dynamic collection of tuples.

\section{Problem Statement}
\label{S3}
Given an uncertain relation $T$  of a dynamic collection of tuples, such that each tuple $t_i \in T$ is associated with a membership probability value $Pr(t_i) > 0$ and a score $score(t_i)$ computed based on a scoring function, the goal is to retrieve the Top-$k$ tuples. 

We use the parameterized ranking function $PRF^e(\alpha)$ proposed by~\cite{R8} in this paper. $PRF^e(\alpha)$ is defined as,

\begin{equation} 
	\label{eq_prfe_def}
	\begin{split}
		\Upsilon(t_i)& = \sum_{r} \alpha^{r-1} \times Pr(t_i, r)
	\end{split} 
\end{equation}

where $\alpha$ is a constant and $Pr(t_i, r)$ denotes the probability of a tuple $t_i$ being ranked at position $r$ across all possible worlds\symbolfootnote[1]{$Pr(t_i, r) = 0$, for $r>i$.}. A top-$k$ query returns the $k$ tuples with highest $\Upsilon$ values. We refer to $\Upsilon(t_i)$ as the {\tt \small rank-score} of tuple $t_i$. In this paper, we adopt the $x$-relation model to capture corrections. An $x$-tuple $\tau$ specifies a set of exclusive tuples, subject to the constraint $\sum_{t_i\in \tau} Pr(t_i) \leq 1$. In a randomly instantiated world $\tau$ takes $t_i$ with probability $Pr(t_i)$, for $i= 1, 2, ...,|\tau|$ or does not appear at all with probability $1 - \sum_{t_i\in \tau} Pr(t_i)$. Here $|\tau|$ represents the number of tuples belonging to set $\tau$. Let $\tau(t_i)$ represents an $x$-tuple to which tuple $t_i$ belongs to. In $x$-relation model, $T$ can be thought of as a collection of pairwise-disjoint $x$-tuples. Let $\sum_{\tau \in T} |\tau| = N$ i.e. there are total $N$ tuples in an  uncertain relation $T$. Without loss of generality, we assume all scores to be unique and let $t_1, t_2, ..., t_N$ denotes ordering of the tuples in $T$ when sorted in descending order of the score $(score(t_i) > score(t_{i+1}))$. From now onwards we represent $Pr(t_i)$ by short notation $p_i$ for simplicity.

\section{Computing $PRF^e(\alpha)$}
\label{S4}
In this section, we derive a closed form expression for the {\tt \small rank-score}  $\Upsilon(t_i)$, followed by an algorithm for retrieving the Top-$1$ tuple from a collection of independent tuples. In the next section we show that this approach can be easily extended to a data structure for efficiently retrieving Top-$k$ tuples from a dynamic collection of tuples. We begin by assuming tuple independence and then consider correlated tuples, where correlations are represented using $x$-tuples.

\subsection{Assuming tuple independence:}

When all tuples are independent, tuple $t_i$ appears at position $r$ in a possible word $pw$ if and only if exactly $(r-1)$ tuples with a higher score value appear in $pw$. Let $S_{i,r}$ be the probability that a randomly generated world from $\{t_1, t_2, ...,t_i\}$ has exactly $r$ tuples. Then, probability of a tuple $t_i$ being ranked at $r$ is given as 

\begin{equation} 
	\label{eq_urank}
	\begin{split}
Pr(t_i, r) = p_i S_{i-1,r-1}
	\end{split} 
\end{equation}

where,
\[ S_{i,r} = \left\{ \begin{array}{ll}
         p_i S_{i-1,r-1} + (1-p_i) S_{i-1,r} & \mbox{if $i \geq r >0$}\\
		 1 & \mbox{if $i = r  =0$}\\
         0 & \mbox{otherwise}.\end{array} \right. \]

Using above recursion for $S_{i,r}$ and equation~\ref{eq_prfe_def}, ~\ref{eq_urank},

\begin{equation*}
\begin{split}
 \Upsilon(t_i) &= \sum_r  \alpha^{r-1} Pr(t_i,r) = \sum_r \alpha^{r-1} p_i S_{i-1,r-1}\\
\frac{\Upsilon(t_i)}{p_i} &=  \sum_r \alpha^{r-1} S_{i-1,r-1} =  \sum_r \alpha^{r} S_{i-1,r}
 \end{split}
\end{equation*}
\\
Similarly,
\begin{equation*}
\begin{split}
\frac{\Upsilon(t_{i+1})}{p_{i+1}} &= \sum_r \alpha^{r} S_{i,r} 
\\ &= \sum_r \alpha^{r}( p_i S_{i-1,r-1}+(1-p_i)S_{i-1,r})
\\&= \alpha p_i \sum_r \alpha^{r-1} S_{i-1,r-1} + (1-p_i)\sum_r \alpha^{r} S_{i-1,r}
\\&= (1-(1-\alpha)p_i)\Upsilon(t_i)/p_i 
 \end{split}
\end{equation*}

We have the base case, $\Upsilon(t_1)=p_1$. Therefore,
\begin{equation}
	\label{eq_prfe_ind}
\begin{split}
 \Upsilon(t_i)&=  p_i \prod_{j<i} (1-(1-\alpha)p_j)
 \end{split}
\end{equation}

Now, we analyze the contribution of a tuple $t_i$ towards global ranking over $T$ using the above formula as follows.
\begin{itemize}
\item Tuple $t_i$ contributes $m_i =  p_i$ for the computation of its own {\tt \small rank-score}.
\item Tuple $t_i$ contributes $c_i = 1-(1-\alpha)p_i$ of computing {\tt \small rank-score} for all tuples having score less than that of $t_i$.
\end{itemize}

\begin{theorem}
When all tuples in $T$ are independent, {\tt \small rank-score} of a tuple $t_i$ can be computed as follows,

\begin{equation*}
\begin{split}
 \Upsilon(t_i)&=  m_i \prod_{j<i} c_j
 \end{split}
\end{equation*}
where $m_i = p_i$ and $c_j = 1-(1-\alpha)p_j$
\end{theorem} 
\qed

\vspace{1.8ex}
\noindent
\emph{Answering Top-$1$ query:}\\
\indent
We use a divide and conquer approach for answering top-$1$ query on $T$, which forms the basis for our data structure in later section. Let the given relation $T=\{t_1, t_2, ..., t_N\}$ be partitioned into sub-reltations $T_l=\{t_1, t_2, ..., t_{\lceil{N/2}\rceil}\}$ and $T_r=\{t_{{\lceil{N/2}\rceil}+1}, t_{{\lceil{N/2}\rceil}+2}, ..., t_N\}$. Also let $t^l$ and $t^r$ represent the top-$1$ answer for $T_l$ and $T_r$ with {\tt rank-scores} $\Upsilon_{T_l}(t^l)$ and  $\Upsilon_{T_r}(t^r)$ respectively, where $\Upsilon_{T_l}(t^l)$ is computed by considering only those tuples $t_j \in T_l$ and $\Upsilon_{T_r}(t^r)$ is is computed by considering only those tuples $t_j \in T_r$.\\

For $t_i \in T_l$,
\begin{equation*}
	\label{eq_ind}
\begin{split}
\Upsilon_{T_l}(t_i)&=  m_i \prod_{\substack{j<i \\ t_j \in T_l}} c_j 
 \end{split}
\end{equation*}

and similarly for $t_i \in T_r$,
\begin{equation*}
\begin{split}
\Upsilon_{T_r}(t_i)&=  m_i \prod_{\substack{j<i \\ t_j \in T_r}} c_j 
 \end{split}
\end{equation*}

Now when both relations $T_l$ and $T_r$ are merged to form $T$, we make the following observations using the above analysis:
\begin{itemize}
\item The contribution of each tuple towards its own {\tt \small rank-score} remains unchanged.
	
\item Since all the tuples in $T_r$ have a lower score value than any tuple $t_i \in T_l$ they do not contribute towards the {\tt \small rank-score} value of $t_i$ computed over entire relation $T$. Thus $\Upsilon(t_i) = \Upsilon_{T_l}(t_i)$. Hence $t^l$ still has the highest {\tt \small rank-score} value $\Upsilon(t^l)$ among the tuples in $T_l$.

\item Since all the tuples in $T_l$ have higher score value than any tuple $t_i \in T_r$, each $t_j \in T_l$ contributes $1-(1-\alpha)p_j$ towards {\tt \small rank-score} value of $t_i$ computed over entire relation $T$. Let $C_l = \prod_{t_j \in T_l} c_j = \prod_{t_j \in T_l} 1-(1-\alpha)p_j$ represents overall contribution of sub-relation $T_l$. Then $\Upsilon(t_i) = C_l \Upsilon_{T_r}(t_i)$. Since {\tt \small rank-score} value of every tuple $t_i \in T_r$ gets scaled by the same factor $C_l$, $t^r$ still has the highest {\tt \small rank-score} value $\Upsilon(t^r)$ among the tuples in $T_r$.
\end{itemize}

Therefore the top-$1$ answer over uncertain relation $T$ can be chosen from $t^l$ and $t^r$ based on the their {\tt \small rank-score} values computed over the entire relation.

\subsection{Supporting correlations}

If $t_i$ has some preceding alternatives, then the event that $t_i$ appears is no longer independent of the event that exactly $j-1$ tuples appear in $\{t_1, t_2, ...,t_{i-1}\}$, as in equation~\ref{eq_urank}. Hence equation~\ref{eq_prfe_ind} cannot be used to compute the {\tt \small rank-score} of a tuple $t_i$.  To overcome this difficulty, we convert the relation $T$ to $\bar{T^i}$ where all the tuples are independent~\cite{R12}. Let $\tau^i = \{t_j| t_j \in \tau, j<i\}$. Now for  each $x$-tuple $\tau \in T$, we create an $x$-tuple $\bar{\tau} = \{\bar{t}\}$ in $\bar{T^i}$, where $p(\bar{t}) = Pr(\tau^i)$ with one exception. For tuple $\bar{t} \in \bar{T^i}$ which corresponds to $\tau(t_i) \in T$, we use $Pr(\bar{t} )= p_i$, where $\tau(t_i)$ is the $x$-tuple to which the tuple $t_i$ belongs to.\\
\\
For example, $T = \{ \tau_1, \tau_2, \tau_3 \}$ where, $\tau_1=\{t_1, t_3, t_6\}, \tau_2 = \{ t_2, t_7\}$ and $\tau_3=\{ t_4, t_5\}$. Then $\tau_1^5 = \{ t_1, t_3\}$ and $\tau(t_5) = \tau_3$.

 This conversion takes into account the fact that only tuples with a  score higher than that of $t_i$ contribute to $Pr(t_i,r)$ as well as to $\Upsilon(t_i)$, and the presence of $t_i$ implies absence of all its related tuples.

Since all the tuples in $\bar{T^i}$ are independent among themselves, we can now use  equation~\ref{eq_prfe_ind} on $\bar{T^i}$ to compute the {\tt \small rank-score} of tuple $t_i$. Combining  related tuples into a representative tuple $\bar t$  does not affect $\Upsilon(t_i)$ here, since the probability that $\bar{t}$ appears is the same as the probability that one tuple in $\tau \in T$ with score higher than $score(t_i)$ appears. Therefore,
\begin{equation}
	\label{eq_prfe_dep}
\begin{split}
 \Upsilon(t_i)&=  p_i \prod_{\substack{\bar{t} \in {\bar{T^i}} \\ \bar{\tau}(\bar{t}) \neq \tau(t_i)}} (1-(1-\alpha)Pr(\bar{t}))\\
&=  p_i \prod_{\substack{\tau \in T \\ \tau \neq \tau(t_i)}} (1-(1-\alpha)Pr(\tau^i))
 \end{split}
\end{equation}

Now, we analyze the contribution of an $x$-tuple towards global ranking over $T$ using the above formula as follows.
\begin{itemize}
\item $x$-tuple $\tau$ contributes $m_i = p_i$ for computing {\tt \small rank-score} of a tuple $t_i \in \tau$.
\item $x$-tuple $\tau$ contributes $c_i = 1-(1-\alpha) Pr(\tau^i)$ for computing {\tt \small rank-score} of a tuple $t_i \notin \tau$.
\end{itemize}

\vspace{1.8ex}
\noindent
\emph{Answering Top-$1$ query:}\\
\indent
Again, we attempt to use a divide and conquer algorithm for answering top-$1$ query on $T$ by partitioning relation $T=\{t_1, t_2, ..., t_N\}$ into sub-relations $T_l=\{t_1, t_2, ..., t_{\lceil{N/2}\rceil}\}$ and $T_r=\{t_{\lceil{N/2}\rceil+1}, t_{\lceil{N/2}\rceil+2}, ..., t_N\}$ and assuming $t^l$ and $t^r$ represent the top-$1$ answers for $T_l$ and $T_r$ respectively. If property that $t^l$ and $t^r$ remains highest {\tt \small rank-score} tuples in their respective sub-relations even after merging of $T_l$ and $T_r$, holds true then reporting top-$1$ for relation $T$ can be done by simply comparing {\tt \small rank-score} values of $t^l$ and $t^r$ over entire relation $T$. Unfortunately, this property may not hold true for $t^r$. 

To illustrate the problem, consider an uncertain relation $T=\{t_1, t_2, t_3, t_4\}$ with $p_1 = 0.35, p_2 = 0.3, p_3 = 0.4, p_4 = 0.45$ and tuples $t_2$ and $t_3$ are mutually exclusive. Using equation~\ref{eq_prfe_dep}, {\tt rank-scores} can be computed as follows ($\alpha = 0.8$):\\
$\Upsilon(t_1)= 0.35$\\
$\Upsilon(t_2)= 0.3(1-0.2\times0.35) = 0.28$\\
$\Upsilon(t_3)= 0.4(1-0.2\times0.35) = 0.37$\\
$\Upsilon(t_4)= 0.45(1-0.2\times0.35)(1-0.2\times(0.3+0.4)) = 0.36$\\

Top-$1$ query on $T$ should return tuple $t_3$ with highest {\tt \small rank-score} value $0.37$. By adopting the divide and conquer approach to tackle the problem, we partition the given relation into $T_l=\{t_1, t_2\}$ and $T_r=\{t_3, t_4\}$. Top-$1$ query is applied to these sub-relations as follows.\\ 
$\Upsilon_{T_l}(t_1)= 0.35$\\ 
$\Upsilon_{T_l}(t_2)= 0.3(1-0.2\times0.35) = 0.28$\\
\\
$\Upsilon_{T_r}(t_3)= 0.4$\\
$\Upsilon_{T_r}(t_4)= 0.45(1-0.2\times0.4) = 0.41$\\

Thus $t_1$ and $t_4$ will be reported from $T_l$ and $T_r$ as top-$1$ answers respectively. By simple merge operation, which computes {\tt \small rank-score} values for $t_1$, $t_4$ over relation $T$ and compares them, $t_1$ will be reported as top-$1$ answer for $T$. However actual top-$1$ answer is tuple $t_3$. The fact that dependance of $t_2$ and $t_3$ was ignored while answering top-$1$ over sub-relation $T_r$ is the root cause behind the disturbance in relative ordering of $t_3$ and $t_4$.

Therefore in order to maintain the relative ordering of tuples based on their {\tt \small rank-score} over entire relation during merge, we redefine the expressions for contributions as follows.
Here we use the notation $\hat{p}_i$ for sum of probabilities of all tuples $t_j$ which are related to $t_i$ and have score greater than the score of $t_i$ (i.e. $j<i$). In the above example $\hat{p}_3= p_2=0.3$. 

\begin{equation*}
\begin{split}
\hat{p}_i = Pr([\tau(t_i)]^i) = \sum_{\substack{\tau(t_i) = \tau(t_j)\\ j < i}} p_j
\end{split}
\end{equation*}

Now equation~\ref{eq_prfe_dep} can be re arranged as follows,

\begin{equation*}
\begin{split}
 \Upsilon(t_i)&=  \frac{p_i}{(1-(1-\alpha)\hat{p}_i)}  \prod_{\tau \in T } (1-(1-\alpha)Pr(\tau^i))  
 \end{split}
\end{equation*}

\begin{equation*}
\begin{split}
 \frac{\Upsilon(t_i)}{m_i}&=   \prod_{\tau \in T } (1-(1-\alpha)Pr(\tau^i))
 \end{split}
\end{equation*}

where $m_i =\frac{p_i}{(1-(1-\alpha)\hat{p}_i)}$\\
\\
similarly,

\begin{equation*}
\begin{split}
 \frac{\Upsilon(t_{i+1})}{m_{i+1}}&=   \prod_{\tau \in T } (1-(1-\alpha)Pr(\tau^{i+1}))
 \end{split}
\end{equation*}

Here note that $Pr(\tau^{i}) = Pr(\tau^{i+1})$ for all $\tau \neq \tau(t_i)$. From the above two equations,

\begin{equation*}
\begin{split}
 \left( \frac{\Upsilon(t_{i+1})}{m_{i+1}}\right)  /\left( \frac{\Upsilon(t_{i})}{m_{i}} \right)&=   \frac{1-(1-\alpha)Pr([\tau(t_i)]^{i+1})}{1-(1-\alpha)Pr([\tau(t_i)]^i)}\\
 &=\frac{1-(1-\alpha)(\hat{p}_i+p_i)}{1-(1-\alpha)\hat{p}_i}\\
 &=c_i
 \end{split}
\end{equation*}

The base case is $\Upsilon(t_1) =p_1$. Therefore we can rewrite  equation~\ref{eq_prfe_dep} as follows, 
\begin{equation}
\begin{split}
\frac{\Upsilon(t_{i+1})}{m_{i+1}}= c_i \frac{\Upsilon(t_{i})}{m_{i}}= c_ic_{i-1}\frac{\Upsilon(t_{i-1})}{m_{i-1}}=...= \prod_{j\leq i} c_j
 \end{split}
\end{equation}

The result is summarized in following theorem. 

\begin{theorem}
For an uncertain relation $T$, {\tt \small rank-score} of a tuple $t_i$ can be computed as,
\begin{equation*}
\begin{split}
 \Upsilon(t_i)&=  m_i \prod_{j<i} c_j
 \end{split}
\end{equation*}

where $m_i =\frac{p_i}{(1-(1-\alpha)\hat{p}_i)}$, $c_i=\frac{1-(1-\alpha)(\hat{p}_i+p_i)}{1-(1-\alpha)\hat{p}_i}$ and  
$\hat{p}_i = \sum t_r$, where $t_i$ and $t_r$ are mutually exclusive and  $r<i$.
\end{theorem} 
\qed

This  equation is applicable for dependent as well as independent tuples. Note that here $m_i$ and $c_i$ are dependent only on the tuples which are related to $t_i$, hence can be computed/updated efficiently. Moreover, the contribution $c_i$ of a tuple $t_i$ to the {\tt \small rank-score} of a tuple $t_j$ is the same for all $j>i$. Hence, the relative ordering will not change even if we use our divide and conquer approach.

Consider the same example as before. We begin by computing values of $m_i$ and $c_i$ for each tuple.\\
\\
$m_1=0.35 \\ m_2 = 0.3 \\ m_3 = 0.4/(1-0.2\times0.3) = 0.43 \\ m_4= 0.45\\ \\
c_1 = (1-0.2\times0.35)=0.93\\
c_2= (1-0.2\times0.3)=0.94\\
c_3=(1-0.2\times(0.3+0.4))/(1-0.2\times0.3)=0.91\\
c_4= (1-0.2\times0.45)=0.91\\$
\\
Now, we partition $T$ into  $T_l=\{t_1, t_2\}$ and $T_r=\{t_3, t_4\}$ and apply Top-$1$ query to these sub-relations.\\  \\
$\Upsilon_{T_l}(t_1)= m_1= 0.35$\\ 
$\Upsilon_{T_l}(t_2)= m_2\times c_1=  0.3\times 0.94 = 0.28$\\
\\
$\Upsilon_{T_r}(t_3)= m_3=0.43$\\
$\Upsilon_{T_r}(t_4)= m_4\times c_3= 0.45\times0.91 = 0.41$

It can be seen that from $t_1$ and $t_3$ are chosen as Top-$1$ from $T_l$ and $T_r$ respectively. During next comparison, $t_3$ ($\Upsilon(t_3)= m_3\times c_1 \times c_2 =0.37$) will be reported as the Top-$1$ tuple, which is correct.

\section{Our Data Structure:}
\label{S5}
In the earlier sections, we derived the simple closed form expression for calculating $ \Upsilon(t_i)$ for a tuple $t_i$. Now our task is to maintain  a dynamic collection of tuples, such that for a given query $k$, we  retrieve Top-$k$  {\tt rank-scored} tuples efficiently.  We use data structural approach for this problem. Our structure is a  balanced binary search tree $\Delta$ such that each leaf corresponds to a tuple in an uncertain relation $T$. Moreover, leaves in the tree are sorted in decreasing order of the score i.e. leaves $\ell_1,\ell_2,...,\ell_N$ of the tree represent tuples $t_1, t_2, ..., t_N$ in the same order from left to right, such that $score(t_i) > score(t_{i+1})$. Let $T_u$ represents the sub-relation containing tuples associated with leaves of a subtree rooted at node $u$. i.e. $T_u = \{t_{u'}, t_{u'+1},..., t_{u''} \}$ and $\ell_{u'}$ represents the left-most and $\ell_{u''}$ represents the right-most leaf of node $u$. At each node $u$, we store a triplet ($top_u,M_u,C_u$) such that:
\begin{itemize}
\item $top_u$ is the tuple (represented by $\ell_{u^*}$) with highest {\tt \small rank-score} among tuples in sub-relation $T_u$. Here ${u'} \leq u^* \leq u''$.
\item $M_u$ is the contribution of all tuples in $T_u$ towards {\tt \small rank-score} of tuple $top_u$. 
\begin{equation*}
\begin{split}
 M_u = m_{u^*}\prod_{ u' \leq i < u^*} c_i
 \end{split}
\end{equation*}
\item $C_u$ is the contribution of all tuples in $T_u$ towards tuple $t_i$ such that $i > u''$, where $\ell_{u''}$ is the right-most leaf of the subtree rooted at node $u$.
\begin{equation*}
\begin{split}
 C_u = \prod_{u' \leq i < u''} c_i
 \end{split}
\end{equation*}
\end{itemize}

Since our data structure stores only a constant number of information at each node, and the number of nodes are bounded by $O(N)$, the total space requirement of our data structure is $O(N)$.

If node $u$ is a leaf node representing the tuple $t_i$, then $M_u=m_i, top_u =t_i $ and $C_u=c_i$. If $u$ is an internal node, this information can be computed using the  {\tt MERGE} operation given below. Figure~\ref{fig:p1} shows an example for the uncertain data in table ~\ref{tab2}.

\begin{figure} [h]
    {\tt MERGE(u)}\\
	\indent \hspace {10pt} $v = left-child(u)$\\
	\indent \hspace {10pt} $w = right-child(u)$\\
    \indent \hspace {10pt} $M_u$ = max $(M_v, C_v \times M_w)$\\
    \indent \hspace {10pt} $top_u=top_v$, if $M_v >C_v \times M_w$, else $top_u=top_w$\\
    \indent \hspace {10pt} $C_u=C_w \times C_w$
\end{figure}

\begin{theorem}
The data structure $\Delta$ maintains a dynamic collections of tuples such that  Top-$1$ tuple, $t^1= top_{root}$ and $\Upsilon(t^1) = M_{root}$.
\end{theorem}

\emph{Proof by contradiction:} Let $t_a$ be the actual Top-$1$ and $top_{root} \neq t_a$. Let $u$ be the closest node from root, such that $top_u = t_a$, that means  $top_{parent(u)} = t_b \neq t_a$. This is because   during the merge operation at $parent(u)$, $ m_a \prod_{x\leq i < a} c_i <   m_b \prod_{x\leq i < b} c_i$ , where $\ell_x$ is the leftmost leaf of $parent(u)$.  Multiplying both the sides of the equation with $\prod_{i<x} c_i$, we get $\Upsilon(t_a) < \Upsilon(t_b)$, which is a contradiction to the statement that $t_a$ is the highest {\tt rank-scored} tuple. Therefore $t^1(=t_a)$ will always be at the root and $M_{root} =  m_a\prod_{ 1 \leq i < a} c_i = \Upsilon(t_a) = \Upsilon(t^1) $. 
\qed
\\

\begin{table}[!t]
\centering

  \caption{Calculation of {\tt rank-scores} (with $\alpha = 0.9$) of tuples in table~\ref{tab_data}: $t_1$ ,$\{t_2, t_4\}$, $\{t_3, t_6\}$, $t_5$ }
      \label{tab2}

    \begin{small}
    \begin{tabular}{|l|l|l|l|l|l|}
    \hline
    Tuple & Prob & $m$ & $c$ & $\Upsilon$ \\ \hline
    $t_1$ & 0.30 & 0.300 & 0.970 & 0.300 \\ \hline 
    $t_2$ & 0.40 & 0.400 & 0.960 & 0.388 \\ \hline 
    $t_3$ & 0.20 & 0.200 & 0.980 & 0.186 \\ \hline 
    $t_4$ & 0.50 & 0.521 & 0.948 & 0.475 \\ \hline 
    $t_5$ & 0.30 & 0.300 & 0.970 & 0.260 \\ \hline 
    $t_6$ & 0.45 & 0.459 & 0.954 & 0.385 \\ \hline 
    \end{tabular}
    \end{small} 
\end{table}

\begin{figure}[h]
\label{fig1}
	\centerline{ \scalebox{0.33}{\includegraphics{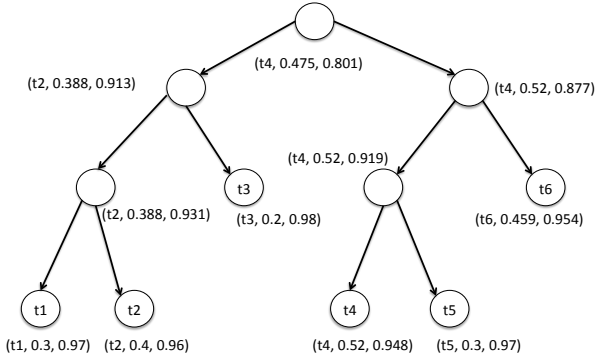}}}
 	\caption{The data structure for uncertain database in Table ~\ref{tab2}}
 	\label{fig:p1}
 \end{figure}

In the following subsections, we show how to perform different operations  such as {\tt update-leaf}, {\tt insert-leaf}  and  {\tt delete-leaf} on this tree. Later, we use these operations for retrieving Top-$k$ tuples, insertion and deletion of tuples.

\subsection{Update-leaf}
The values $m_i$ and $c_i$ within a leaf node $\ell_i$ can be changed in constant time. But this will change the $m$ and $c$ values at all nodes which are in that path from $\ell_i$ to root. Therefore we need to perform {\tt MERGE} operation on all nodes in the path from $\ell_i$ to root, starting from $parent(\ell_i)$. Since the height of a balanced binary tree is bounded by $O(\log N)$, the total  time for {\tt update-leaf}  can also be bounded by $O(\log N)$. 
\begin{theorem}
The $m_i$ and $c_i$ values of a leaf can be updated in $O(\log N)$ time.
\end{theorem}

\subsection{Insert-leaf and delete-leaf}
We first explain, how one-one correspondence between tree leaves and tuples in relation $T$ can be maintained during insertion or deletion of a leaf.
\begin{itemize}
	\item \emph{Insert:} To insert a new leaf, we begin by carrying out standard insert procedure of a binary search tree, which would create a new leaf node $v$. Let $w$ be the parent of this newly created node. Node $w$ being the leaf prior to insertion of $v$, represents a single tuple from $T$ and should remain as a  leaf after insertion of $v$ as well. This can be achieved by creating a new internal node $u$, which becomes the parent of $v$ and $w$.
	
	\item \emph{Delete:} If deletion of a node results in an internal node with only one child, we perform recursive delete on  that internal node.
\end{itemize}	

After insert or delete of a leaf node $\ell_i$, we need to update the $M$ and $C$ values at each node along the path of insertion or deletion. This can be achieved by performing {\tt MERGE} operation in bottom-up fashion beginning with $parent(\ell_i)$. If tree goes out of balance after insert or delete, necessary rebalancing may force further re-computation at nodes whose left or right subtree is changed. However, such nodes are bounded by the height ($O(\log N)$) of the tree. Hence {\tt Insert-leaf} and {\tt leaf-delete} operations can be done $O(\log N)$ time.


\subsection{Retrieving Top-$k$ tuples}
In theorem 3, we proved that, by {\tt MERGE} operation the Top-$1$ tuple $t^1$  will be the propagated to root node as $top_{root}$. Therefore $t^1$  can be retrieved in constant time. In order to retrieve the Top-$2$ tuple $t^2$, we use the following strategy. After retrieving $t^1$, we set $\Upsilon(t^1)=0$.  As a result, the next highest  ${\tt rank-scored}$ tuple $t^2$ will be propagated as $top_{root}$ instead of $t^1$. This can be achieved by performing {\tt Update-leaf} operation on leaf $\ell_j$ (leaf representing the current $top_{root} = t_j$), with it $m_j$ value set to zero. As $c_j$ remains unchanged, update operation affects only the computation of {\tt rank-score} of $t_j$ leaving {\tt rank-score} of all other tuples unchanged. Repeating the same process, we can retrieve top-$k$ tuples with highest {\tt rank-score} values. We can revert back the changes done in data structure for answering top-$k$ query by restoring the $m$ values for $k$ retrieved tuples using {\tt Update-leaf} operation.

\begin{figure} [h]
{\tt Top-$k$}\\ 
  \indent \hspace {10pt} for $i = 1$ to $k$\\
	\indent \hspace {20pt} $t_j$ = $top_{root}$\\
  \indent \hspace {20pt}	report $top_{root}$ as top-$i$ tuple\\
\indent \hspace {20pt}  	{\tt Update-leaf}($t_j$) with $m_j=0$\\
\end{figure}

\begin{figure}[h]
	\centerline{ \scalebox{0.33}{\includegraphics{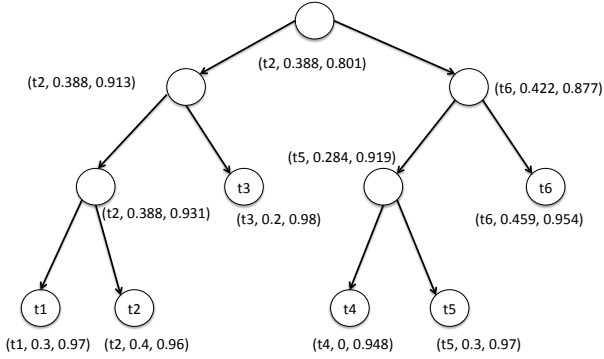}}}
 	\caption{The data structure after setting $m_4 = 0$ for retrieving Top-$2$}
 	\label{fig:p2}
 \end{figure}

Figure~\ref{fig:p2} shows an  example for retrieving Top-$2$ tuple from the uncertain data in table~\ref{tab_data}.
\vspace{1.8ex}
\begin{theorem}
Top-$k$ {\tt rank-scored} tuples can be retrieved  in $O(k\log N)$ time.
\end{theorem}

\emph{Proof:} For every tuple $t_j$ retrieved for answering top-$k$ query, we perform {\tt Update-leaf} operation twice: once for setting $m_j = 0$ so that tuple with next highest {\tt rank-score} can be retrieved and next after reporting top-$k$ answers so as to restore the tree changes. Since {\tt Update-leaf} is a $O(\log N)$ time operation, total time for Top-$k$ retrieval can be bounded by $O(k\log N)$.

\subsection{Insert-tuple and delete-tuple}

Whenever a tuple $t_i$ gets inserted(deleted) from relation $T$, we modify our data structure as follows:
\begin{itemize}
\item We begin by carrying out {\tt Insert-leaf} or {\tt leaf-delete} operation as necessary. If $t_i$ is an independent tuple then at this point all nodes in the tree $\Delta$ have correct values for $C$ and $M$. Hence no further action is necessary.
\item If $t_i$ is not independent, then its insertion(deletion) will change $m_j$ and $c_j$ values for all leaf nodes corresponding to tuple $t_j$ such that $j >i$ and $\tau(t_i) = \tau(t_j)$. These change can be accommodated by performing {\tt Update-leaf} operation on each $\ell_j$.
\end{itemize}

 \begin{figure}[h]
	\centerline{ \scalebox{0.33}{\includegraphics{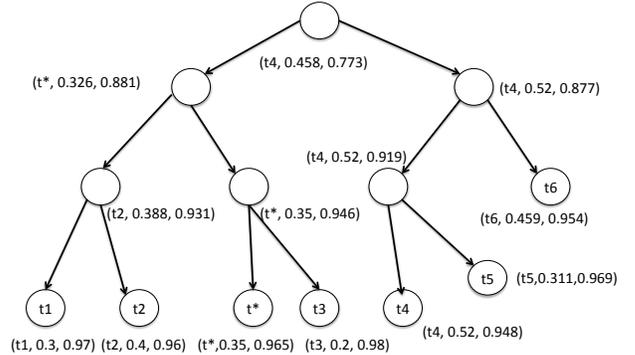}}}
 	\caption{The data structure in fig1 after inserting t* }
 	\label{fig:p3}
 \end{figure}
  
\begin{figure}[h]
	\centerline{ \scalebox{0.33}{\includegraphics{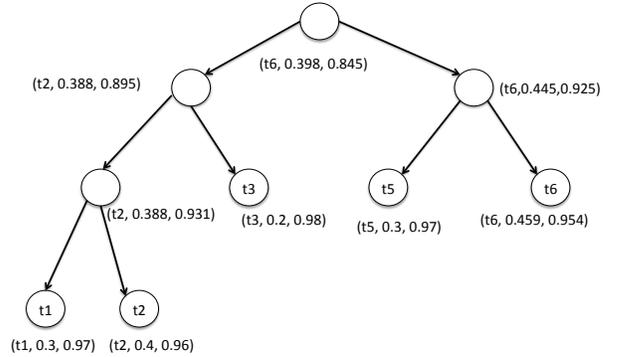}}}
 	\caption{The data structure in fig1 after deleting $t_4$}
 	\label{fig:p4}
 \end{figure}

  Figure~\ref{fig:p3} shows an  example of inserting a new tuple $t^*$(with $score(t_2) > score(t^*) > score(t_3)$) and is mutually exclusive with $t_5$  in the uncertain data in table~\ref{tab2} and figure~\ref{fig:p4} shows an example for deletion of a tuple.
  
Thus insertion(deletion) of a tuple can result in one {\tt Insert-leaf} or {\tt leaf-delete} operation and at max $|\tau(t_i)|$  {\tt Update-leaf} operations. Since any $x$-tuple can have only constant number of operations, tuple insertion and deletion can be handled in $O(\log N)$ time. We note that updating of tuples  can be simulated by first deleting and then reinserting it with updated values.

We summarize the space requirement and performance of the proposed data structure in the  following theorem.
\begin{theorem}
A collection of uncertain data can be maintained using a linear size dynamic data structure, which can retrieve Top-$k$ {\tt rank-scored} tuples in $O(k\log N)$ time, and can support insertion or deletion of a tuple $t$  in $O(d\log N)$ time, where $d$ is the number of tuples which are related to $t$. 
\end{theorem} \qed

\section {Experimental Study}
\label{S6}
In this section, we present an experimental study with both synthetic and real data evaluating effectiveness of the data structure in handling changes in underlying database and answering top-$k$ queries. All experiments were conducted on 2.4 GHz Intel Core 2 Duo machine with 2GB memory running MAC OS 10.6.4. 
\vspace{1.5ex}
\\
\noindent \emph{Datasets:} We created a synthetic dataset containing 1,00,000 tuples. Score of a each tuple is chosen uniformly at random from [0,100000] and it's probability is uniformly distributed in ($0.5 \times 10^{-5}, 1.5 \times 10^{-5}$). The number of tuples involved in each $x$-tuple  follows the uniform distribution (2,10). 

\begin{figure}[h]
	\centerline{ \scalebox{0.32}{\includegraphics{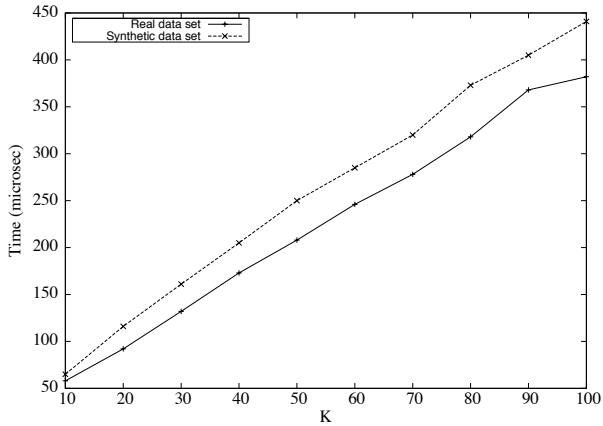}}}
	\caption{Top-$k$ query performance on real and synthetic data}
	\label{varyk}
\end{figure}

Along with synthetic datasets, we also use International Ice Patrol(IIP) Iceberg Sighting Database~\footnote{http://nsidc.org/data/g00807.html}. Each sighting record in the database contains date, location, number of days the iceberg has drifted, etc. As it is crucial to detect the icebergs drifting for long periods, we use the \emph{number of days drifted} as ranking score. The sighting record is also contains a confidence-level attribute according to the source of sighting: R/V (radar and visual), VIS (visual only), RAD (radar only), SAT-LOW (low earth orbit satellite), SAT-MED (medium earth orbit satellite), SAT-HIGH (high earth orbit satellite), and EST (estimated). We converted these seven confidence levels into probabilities 0.8, 0.7, 0.6, 0.5, 0.4, 0.3, and 0.4 respectively. We gathered all records from 1981 to 1991 and 1998 to 2004. Based on it then we created 1,00,000 tuples dataset by repeatedly selecting records randomly.
\vspace{1.5ex}
\\
\noindent \emph{Results:}
For all of our experiments we choose $\alpha = 1-0.9^{50}$. We begin by evaluating the query performance of the data structure. We retrieve top-$k$ tuples from both the datasets for $k$ ranging from 10 to 100. Linear dependance of query time as obtained in the time bounds is evident from the results show in Figure~\ref{varyk}. Also we can note that, correlations among tuples does not affect the query time of our data structure.

\begin{figure}[h]
	\centerline{ \scalebox{0.32}{\includegraphics{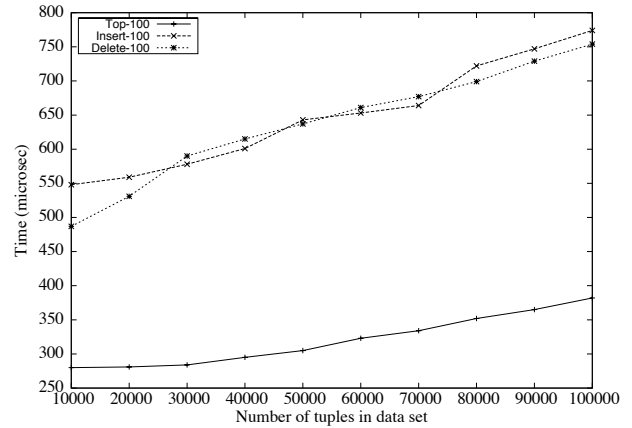}}}
	\caption{Processing (insert, delete, top-$k$) cost on real dataset}
	\label{real}
\end{figure}

\begin{figure}[h]
	\centerline{ \scalebox{0.32}{\includegraphics{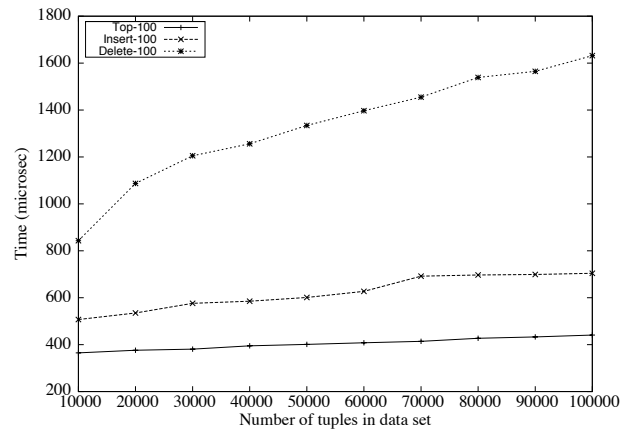}}}
	\caption{Processing (insert, delete, top-$k$) cost on synthetic dataset}
	\label{syn}
\end{figure}

Next set of experiments conducted shows efficiency of our data structure in handling tuple insertions and deletions. Time required for inserting and deleting 100 tuples is measured for datasets of varying sizes. Figure~\ref{real} and ~\ref{syn} shows that processing time per tuple increase slowly with data size. Whenever a tuple is inserted or deleted, to maintain the correctness of data structure, we also need to update information for leaves corresponding to its related tuples. As all tuples in real data set are assumed to be independent average insertion/deletion time of a tuple is less than in case of synthetic data having correlations. This can be seen from the results in figure~\ref{real} and ~\ref{syn}. For synthetic dataset, we insert a tuple in dataset such that it is related to existing tuples. We ensure the $x$-tuple probability to be less than 1 to which new tuple is inserted. For deletion, victim tuple is selected at random. Figure~\ref{real} and ~\ref{syn} also shows the effect of varying data size on query performance of data structure.  

Data structure proposed in this paper can be used when data arrives in streaming fashion. Jin et al.~\cite{R5} have studied the problem of answering top-$k$ queries on sliding windows. Our data structure achieves performance comparable to synopses proposed by them in terms of handling tuple insertion and deletions. Even though our data structure takes linear size as compared to these space efficient synopses, it can be noted that they rely on random order stream model used in streams algorithm community~\cite{R13,R14,R16} and in worst case would take linear size as well.

\section{Related Work}
\label{S7}
Uncertain data management has attracted a lot of attention in recent years due to an increase in the number of application domains that naturally generate uncertain data.  These include sensor networks~\cite{R17}, data cleaning~\cite{R18} and data integration~\cite{R19,R20} . Several probabilistic data models have been proposed to capture data uncertainty (e.g TRIO~\cite{R11}, MYSTIQ~\cite{R3}, MayBMS~\cite{R6}, ORION~\cite{R1}, PrDB~\cite{R9}). Virtually all models have adopted  possible worlds semantics. Each data model captures  tuple uncertainty (existence probabilities are attached to the tuples of the database), or attribute uncertainty (probability distributions are attached to the attributes) or both. Further distinction can be made among these models based on support for correlations. Most of the work in probabilistic databases has either assumed independence or supports restricted correlations, mutual exclusion being the most common. Recently proposed approaches~\cite{R9,R7} extend the support for any arbitrary correlations. 

Efforts have been made in recent times to extend the semantics of ``top-$k$'' to uncertain databases. Soliman et al.~\cite{R10} defined the problem of ranking over uncertain databases. They proposed two ranking functions, namely {\tt U-Top$k$} and {\tt U-$k$Ranks}, and proposed algorithms for each of them. Improved algorithms for the same ranking functions were presented later by Yi et al.~\cite{R12}.  Hua et al.~\cite{R4} proposed another top-$k$ definition {\tt PT-$k$} (\emph{probabilistic threshold queries}) and proposed efficient solutions. Cormode et al.~\cite{R2} defined number of key properties satisfied by ``top-$k$" over deterministic data including {\tt exact-$k$, containment, unique-rank, value-invariance,} and {\tt stability}. With each of the existing top-$k$ definition lacking one or more of these properties, Cormode at al.~\cite{R2} proposed yet another ranking function {\tt expected-rank}. As the list of top-$k$ definitions continued to grow, Li et al.~\cite{R8} argued that a single specific ranking function may not be appropriate to rank different uncertain databases and empirically illustrated the diverse, conflicting nature of  parameterized ranking functions that generalize or can approximate many know ranking functions. 

With most of the work for top-$k$ query processing being focused on ``one-shot'' top-$k$ query for static uncertain data, Chen and Yi~\cite{R15} was the first to address the dynamic aspect of uncertain data. They proposed a fully dynamic data structure to support arbitrary insertions and deletions. For an uncertain relation with $N$ tuples, the structure of~\cite{R15}  answers top-$k$ queries in $O(k + \log N)$ time, handles an update in $O(k \log k \log N)$ time and takes $O(N)$ space. However, this structure is tied to a single ranking function i.e. {\tt U-Top$k$} and works only for independent tuples. Moreover, it can be built for some fixed $k$ value and cannot answer a top-$j$ for $j > k$. Dependance of time, required for handling update, on $k$ is also not desirable. Recently, Jin et al.~\cite{R5} proposed a framework for sliding window top-$k$ queries on uncertain streams supporting several ranking functions. This framework assumes {\tt random-order stream model} (tuples arrive in a random order) which significantly reduces the space requirement as compared to the worst-case scenario in which any data structure will have to remember every tuple in the current window.

\section{Conclusions}

In this paper we present a dynamic data structure, which can retrieve top-$k$ tuples in $O(k\log N)$ time and has update cost of $O(\log N)$. We also evaluate efficiency of proposed data structure with experiments using synthetic and real data. It is an open question if, we can improve the top-$k$ retrieval time to $O(k + \log N)$ without sacrificing update time or is there any lower bound for this problem?

\bibliographystyle{IEEEtran}
\bibliography{IEEEabrv,IEEEexample}

\end{document}